\documentclass[letterpaper,twoside,twocolumn,american,english,prl]{revtex4-1}
\usepackage[T1]{fontenc}
\usepackage[latin9]{inputenc}
\usepackage{amsmath}
\usepackage{amssymb}
\usepackage{graphicx}

\makeatletter

\@ifundefined{textcolor}{}
{%
 \definecolor{BLACK}{gray}{0}
 \definecolor{WHITE}{gray}{1}
 \definecolor{RED}{rgb}{1,0,0}
 \definecolor{GREEN}{rgb}{0,1,0}
 \definecolor{BLUE}{rgb}{0,0,1}
 \definecolor{CYAN}{cmyk}{1,0,0,0}
 \definecolor{MAGENTA}{cmyk}{0,1,0,0}
 \definecolor{YELLOW}{cmyk}{0,0,1,0}
}

\usepackage{babel}

\makeatother

\usepackage{babel}
\begin{document}

\title{Renormalization Group Solution of the Chutes\&Ladder Model}

\author{Lauren A. Ball, Alfred C. K. Farris, and Stefan~Boettcher}

\email{sboettc@emory.edu}

\homepage{http://www.physics.emory.edu/faculty/boettcher/}

\selectlanguage{english}%

\affiliation{Dept.~of Physics, Emory University, Atlanta, GA 30322; USA}
\begin{abstract}
We analyze a semi-infinite one-dimensional random walk process with
a biased motion that is incremental in one direction and long-range
in the other. On a network with a fixed hierarchy of long-range jumps,
we find with exact renormalization group calculations that there is
a dynamical transition between a localized adsorption phase and an
anomalous diffusion phase in which the mean-square displacement exponent
depends non-universally on the Bernoulli coin. We relate these results
to similar findings of unconventional phase behavior in hierarchical
networks.
\end{abstract}
\maketitle

\section{Introduction\label{sec:Introduction}}

The variety of real networks found in biology, engineering, social
sciences, and communication provides a need for new ideas to explore
and classify the full range of critical phenomena that emerge as a
result of the complex geometry \cite{Barabasi03,Boccaletti06,Dorogovtsev08,barthelemy_spatial_2010}.
Networks with a hierarchical organization of its sites have a long
history in providing solvable models of statistical systems \cite{Dhar77,Berker79}.
More recently, such networks, when turned hyperbolic with the addition
of small-world links, have received considerable attention due to
a variety of synthetic phase transitions that can be observed in such
structures for well-known equilibrium models such as percolation \cite{Berker09,Boettcher09c,Hasegawa10b}
and Ising ferromagnets \cite{PhysRevLett.108.255703,Boettcher10c,Andrade09,Hinczewski06,Hinczewski07}.
For instance, hyperbolic networks interwoven with geometric backbones
provide solvable examples of discontinuous (``explosive'') percolation
transitions \cite{Boettcher11d,Singh14,Cho14}. Previous studies of
symmetric walks on such networks give simple examples of super-diffusion
and shown close connections with L\'evy flights \cite{SWN,SWlong}.

Here, we consider a strongly biased variant of the familiar persistent
random work \cite{Renshaw94,Weiss94}; an incrementally progressing
walker in the forward direction undertakes back-jumps with a tunable
frequency that is inversely related to the length of the long jump.
But unlike those persistent walks that typically remain within the
universality class of ordinary diffusion \cite{Weiss94}, the asymptotic
behavior of our walks exhibit anomalous diffusion behavior with exponents
that depend on the long-jump bias of the Bernoulli coin $p$. For
instance, for the exponent that relates typical length and time-scales,
$T\sim L^{d_{w}}$, such as determined through the mean-square displacement,
we find
\begin{equation}
d_{w}\left(p\right)=\log_{2}\left(\frac{2-3p}{1-2p}\right),\label{eq:dw}
\end{equation}
which continuously ranges through $0<p\leq\frac{1}{2}$, from straight
ballistic motion, $d_{w}\left(0\right)=1$, to all forms of anomalous
super and sub-diffusion down to complete confinement or adsorption
\cite{PhysRevLett.74.2410}, $d_{w}=\infty$, for $p>\frac{1}{2}$.
For instance, according to Eq. (\ref{eq:dw}), exactly at $p=\frac{2}{5}$
the balance between step-by-step progression and occasional fall-backs
results in an effectively diffusive spreading, $d_{w}=2$, at long
ranges. Among other benefits, the tunability of the spreading walk
provides strong control over transport properties\cite{TASEP09,Boettcher11b}
as well as mixing times. Combined with some forms of control, persistent
walks such as this, for example, have received some renewed attention
recently as a means to accelerate Markov chain algorithms via ``lifting''
\cite{Chen99,Hayes10,Hukushima13}. Furthermore, quantized version
of such walks are a fundamental ingredient in quantum algorithms \cite{Gro97a,AAKV01,Boettcher13a}. 

First and foremost, our study here serves as a simple illustration
of the unusual -- and often non-universal -- scaling behavior for
dynamic processes on complex networks. In particular, hyperbolic networks
such as the one considered here have been shown to possess a number
of interesting, \emph{synthetic} phase transitions in which the scaling
behavior can be controlled by global parameters \cite{Boettcher10c,BoBr12,Singh14b}.
However, our model also provides a sense of what might happen in an
ordinary, one-dimensional lattice with an incremental bias to walk
one direction and back-jumps in the opposite direction, drawn randomly
(annealed) from a L\'evy-flight distribution \cite{Shlesinger93}.
This could be extended, for example, to model the behavior of directional
transport of kinesin \cite{Huang11}, interrupted with finite failure
rate that leads to dissociation off the actin filament to reset the
process. Certain forms of foraging behavior have also been described
in these terms \cite{Viswanathan99,Faustino07,PhysRevE.86.061102,Faustino12}. 

Our discussion is organized as follows. In the next section, we describe
the network we are using and the random walk on it. In Sec. \ref{sec:Numerical-Simulations},
we describe the results of our numerical simulations. In Sec. \ref{sec:Renormalization-Group-Analysis}
we discuss our renormalization group calculation. Finally, we conclude
our discussion in Sec. \ref{sec:Conclusions}.

\section{Network Design\label{sec:Network-Design}}

\begin{figure}
\includegraphics[bb=14bp 14bp 822bp 528bp,clip,width=1\columnwidth]{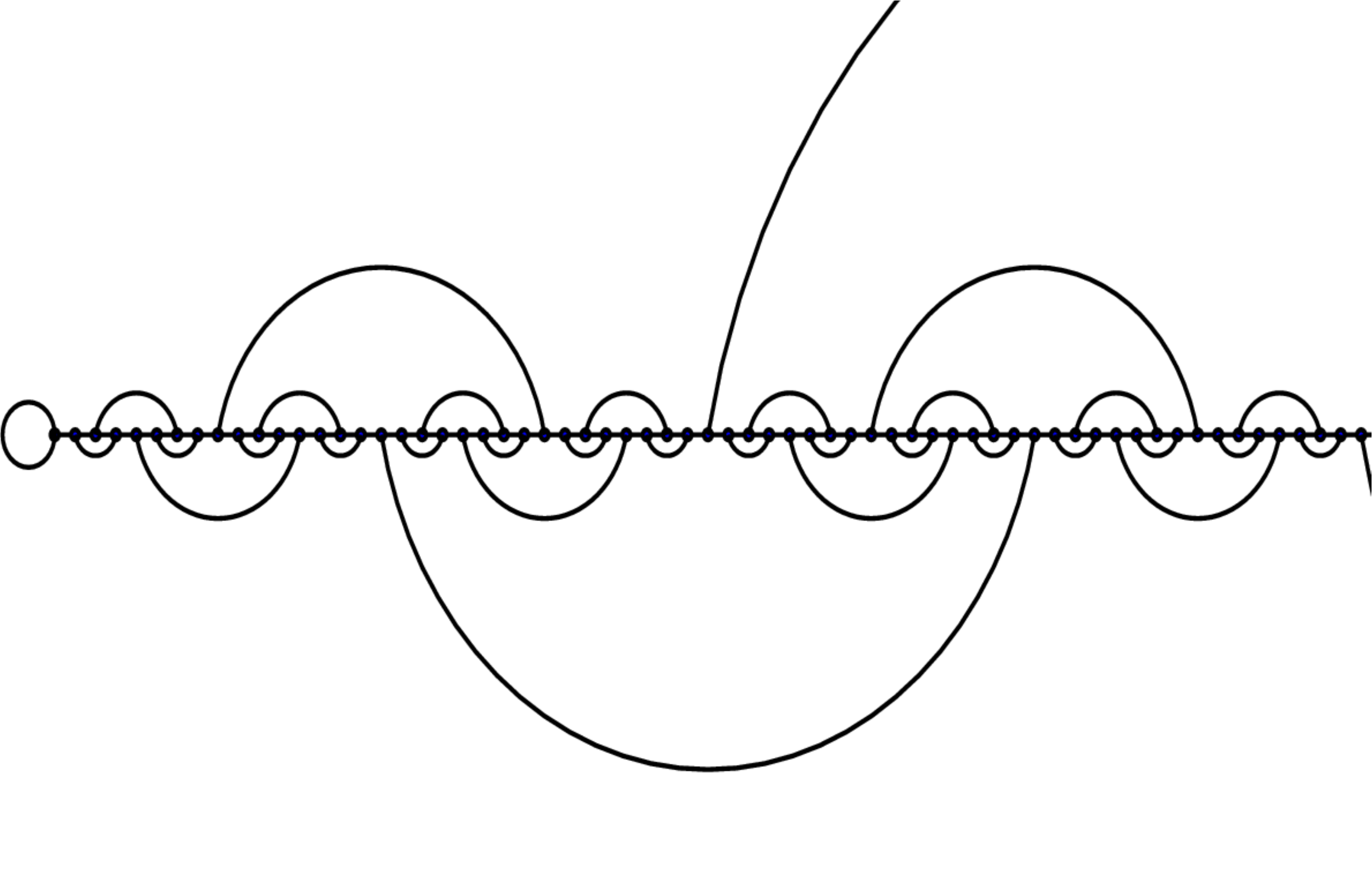}

\protect\caption{\label{fig:3hanoi}Depiction of HN3 on a semi-infinite line. The leftmost
site here is $n=0$, which requires special treatment. }
\end{figure}

The network we are discussing in this paper \cite{SWPRL} consists
of a simple geometric backbone, a one-dimensional line of $N=2^{l}$
sites $\left(0\leq n\leq2^{l},\, l\to\infty\right)$. Each site on
the one-dimensional lattice backbone is connected to its nearest neighbor.
To generate the small-world hierarchy in these graphs, consider parameterizing
any integer $n$ (except for zero) \emph{uniquely} in terms of two
other integers $(i,j)$, $i\geq0$,
\begin{eqnarray}
n & = & 2^{i}\left(2j+1\right).\label{eq:numbering}
\end{eqnarray}
Here, $i$ denotes the level in the hierarchy whereas $j$ labels
consecutive sites within each hierarchy. For instance, $i=0$ refers
to all odd integers, $i=1$ to all integers once divisible by 2 (i.
e., 2, 6, 10,...), and so on. In these networks, aside from the backbone,
each site is also connected with (one or both) of its nearest neighbors
\emph{within} the hierarchy. We obtain the 3-regular network HN3 by
connecting first all nearest neighbors along the backbone, but in
addition also 1 to 3, 5 to 7, 9 to 11, etc, for $i=0$, next 2 to
6, 10 to 14, etc, for $i=1$, and 4 to 12, 20 to 28, etc, for $i=2$,
and so on, as depicted in Fig. \ref{fig:3hanoi}. The site with index
zero, not being covered by Eq. (\ref{eq:numbering}), is clearly a
special place on the boundary of the HN3 that provides an impenetrable
wall and ensures that the walks remains semi-infinite. 

\section{Chutes \& Ladder Model\label{sec:Chutes=000026Ladder-Model}}

\begin{figure}
\centering \includegraphics[bb=60bp 220bp 550bp 600bp,clip,width=1\columnwidth]{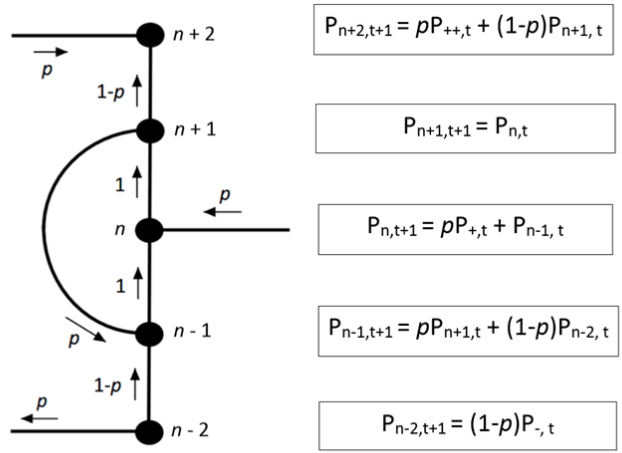}
\protect\caption{\label{fig:4site}Hopping propabilities between sites in the typical
sub-graph of HN3, the network depicted in Fig. \ref{fig:3hanoi}.
Shown are the hopping probabilities from each site at time $t$, and
the corresponding equations describing the probability of the site
being occupied at time $t+1$. (Subscripts ``+'', ``++'', and
``-'' refer to unspecified distant sites connected via long-range
jumps.)}
\end{figure}

We will study biased random walks on HN3. Each site of HN3 has degree
three, i.e., the walker has three possible ways to exit her current
position, as indicated in Fig. \ref{fig:4site}: up or down the backbone,
or along the long-range link. The option with a walk that is unbiased
for hops along the backbone has been studied in Refs. \cite{SWN,SWlong}.
Here, we consider the extreme case of a walker that can move along
the backbone in only one direction (``up'') while along the small-world
links it can only move in the opposite direction (``down''). At
a site at which the long-range link is oriented in the up-direction,
the walker \emph{deterministically} moves up along the backbone. Only
at sites with a long-range link in the down-direction does the walker
have a choice. Thus, all walks are controlled by the parameter $p$,
the Bernoulli coin, which is the probability of a walker to jump off
the lattice in a long-range jump down towards the origin. At such
a site, the walker will move up along the backbone with probability
$(1-p)$ . These probabilities are displayed in Fig. \ref{fig:4site}.
If $p$ were set to zero, we would obtain a ballistic one-dimensional
walk, exclusively biased in the up-direction. For any finite $p$,
the walk resembles the popular children's game of ``Chutes \& Ladders''
\footnote{http://uncyclopedia.wikia.com/wiki/Chutes\_and\_Ladders%
}.

\section{Numerical Simulations\label{sec:Numerical-Simulations}}

In our simulations, each walk begins at site $n=1$ at time $t=0$.
This excludes site $n=0$ from the process, as it can not be reached.
It results in a semi-infinite walk restricted to all sites of positive
index $n$. Having arrived at any site $n$, the procedure for determining
the walker\textquoteright s next step involves first calculating each
path that he could potentially take from its current position, based
on the HN3 geometry. First, we determine the level in the hierarchy
of the possible long-range jump, $i$ in Eq. \eqref{eq:numbering},
by counting how many times the current site index $n$ is divisible
by two. After peeling this factor of $2^{i}$ off $n$, we obtain
$n/2^{i}=2j+1$, an odd number. If now $j$ is also an odd number,
the walker is located at a long-range link in the down-direction;
if instead $j$ is even, the long-range link emanating from $n$ is
oriented in the up-direction and blocked for the walker. In the latter
case, the walker must move up by one step along the backbone. In the
former case, it has two possibilities: with probability $p$ the long-range
link in the down-direction is taken, while with probability $1-p$
the walker will move up along the backbone. 

\begin{figure*}
\includegraphics[bb=0bp 0bp 750bp 550bp,clip,width=0.5\textwidth]{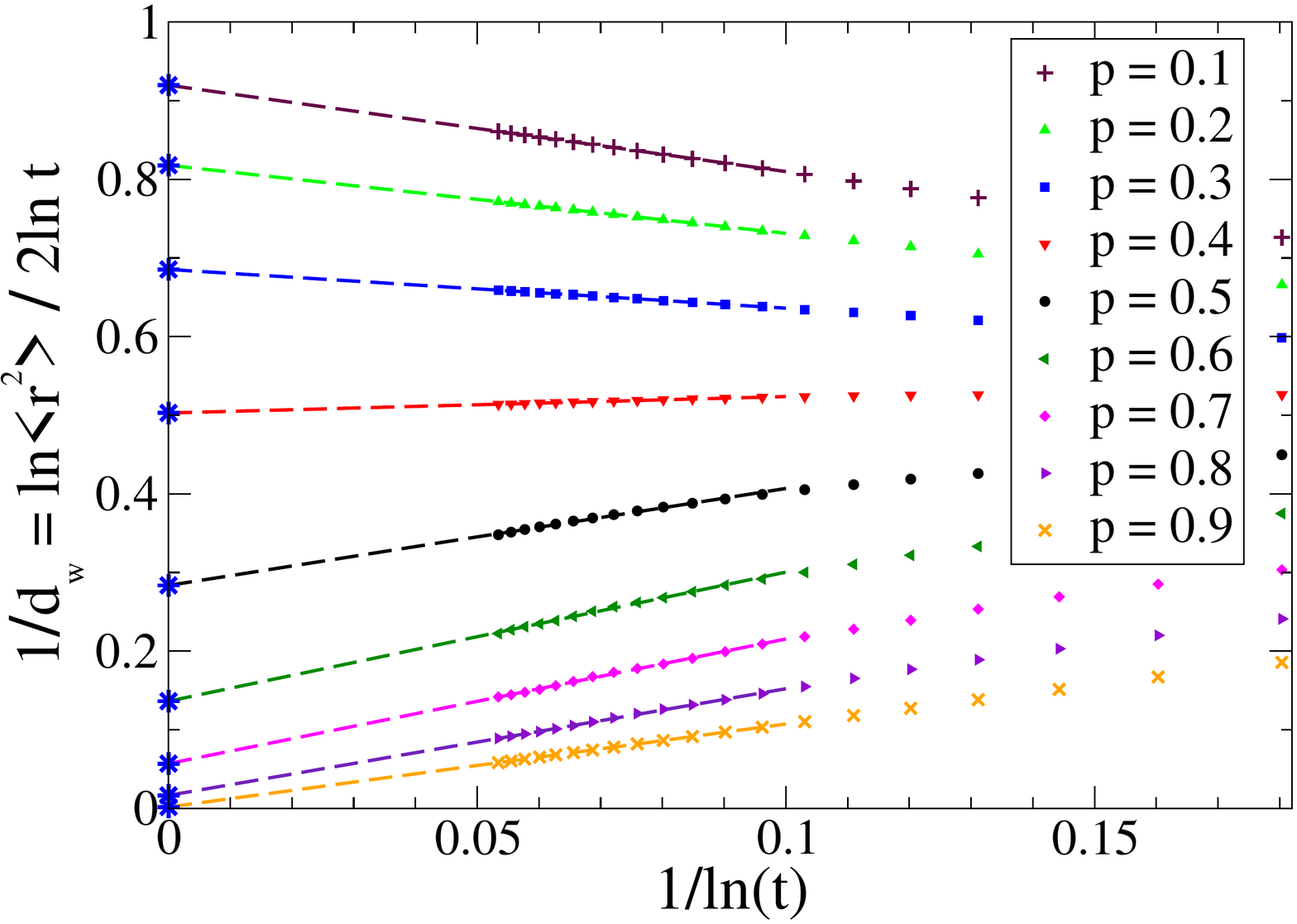}\hfill{}\includegraphics[bb=0bp 0bp 750bp 550bp,clip,width=0.5\textwidth]{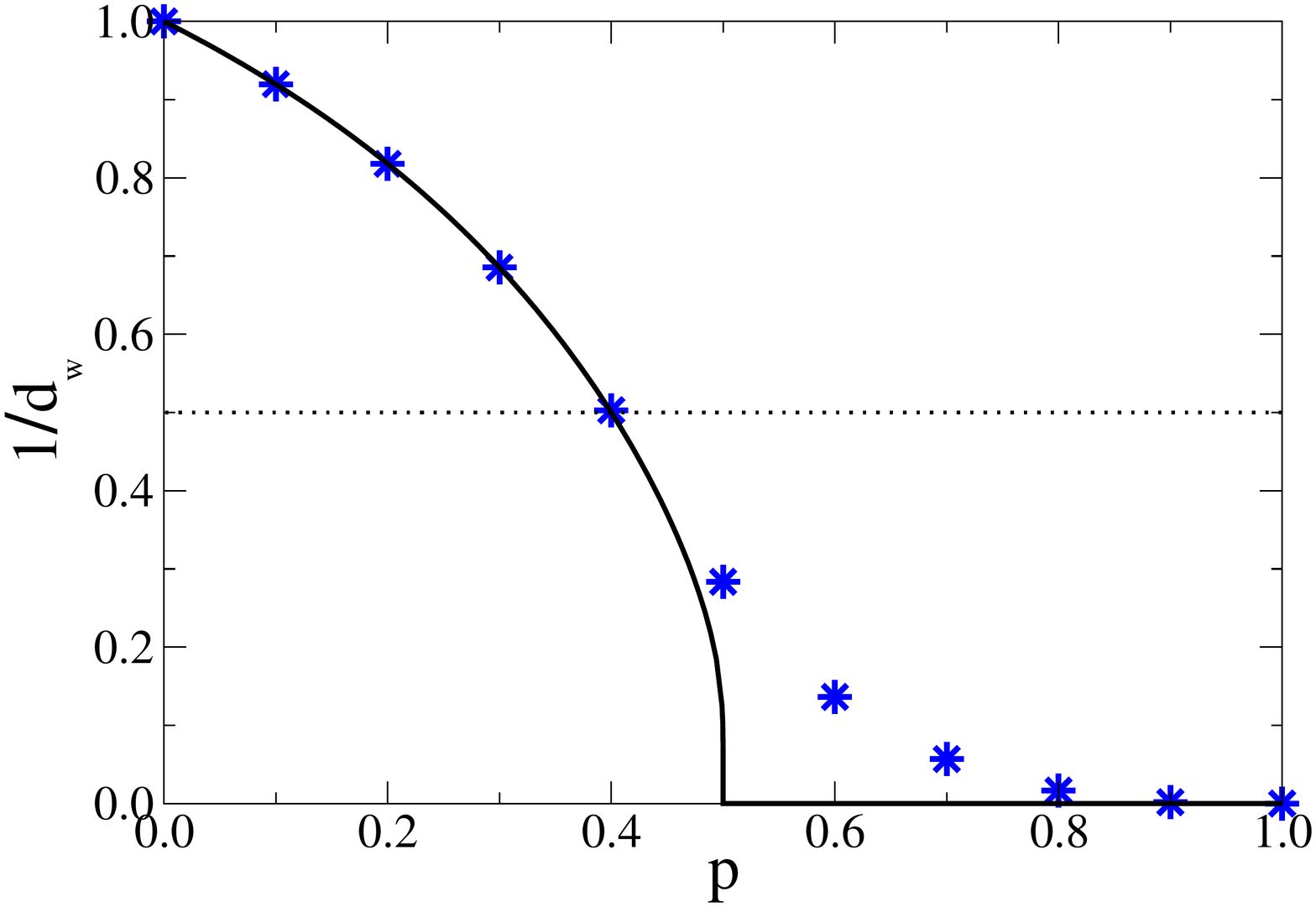}\protect\caption{\label{fig:Extrapolation}(a) Extrapolation of the data obtained for
the mean-square displacement for the walk at different values of the
down-jump probability $p$ from simulations run up to a temporal cutoff
at $t=2^{27}$. We used a linear fit according to Eq. (\ref{eq:Extrapolation})
to the data set for each value of $p$ deep in the asymptotic regime
for large $t$, as indicated by the fitted lines, data for all smaller
$t$ were ignored. While statistical errors are much below the symbol
size, large corrections to linear behavior are apparent for larger
$p$, suggesting large systematic errors. The fitted values for $1/d_{w}$
of the extrapolation for $t\to\infty$ are marked as blue stars at
the intercept. (b) These extrapolated values (blue stars) for the
exponent $d_{w}$ found are plotted as a function of $p$. The solid
line corresponds to the exact RG-result according to Eq. \eqref{eq:dw}. }
\end{figure*}

We have conducted extensive sampling of such a walk for each value
of $p=0.1,0.2,\ldots0.9$ by averaging the mean-square displacement
(MSD) as a function of time over at least $10^{6}$ walks that were
terminated after $t_{{\rm max}}=2^{27}\approx10^{8}$ update steps.
In the expectation of an anomalous power-law relation of the MSD with
time,
\begin{equation}
\left\langle r^{2}\right\rangle \sim D\, t^{\frac{2}{d_{w}}}\label{eq:MSD}
\end{equation}
with some positive constant $D$, in Fig. \ref{fig:Extrapolation}(a)
we have plotted our data in the form of an extrapolation plot, 
\begin{equation}
\frac{\ln\left\langle r^{2}\right\rangle }{2\ln t}\sim\frac{1}{d_{w}}+\frac{\ln D}{\ln t}.\label{eq:Extrapolation}
\end{equation}
When plotted on an $1/\ln t$-scale as the $x$-axis, the data should
extrapolate \emph{linearly} to the desired MSD exponent, $1/d_{w}$,
read off on the intercept with the $y$-axis for $t\to\infty$. In
the figure, we have marked those intercepts with blue stars for each
value of $p$. In Fig. \ref{fig:Extrapolation}(b) we plot those extrapolants
as a function of $p$. In that panel, we have also indicated the exact
RG-result advertised in Eq. (\ref{eq:dw}) as a solid line. Our data
matches that line within errors for all $p<\frac{1}{2}$, while for
values of $p\geq\frac{1}{2}$, where the walk develops a strong bias
towards returning to the origin (i.e., $d_{w}\to\infty$), there are
strong deviations due to slow convergence and, for $p>\frac{1}{2}$,
a break-down of the power-law assumption (i.e., a lack of a linear
extrapolation) in Eq. (\ref{eq:MSD}), as is also apparent from the
break-down of Eq. (\ref{eq:dw}) for $p>\frac{1}{2}$. In that regime,
the walker remains entirely confined near the origin.

The failure of our extrapolation near $p=\frac{1}{2}$ is easy to
understand: For $p\nearrow\frac{1}{2}$, $d_{w}\sim\log_{2}\left(\frac{1}{2}-p\right)$
diverges rapidly, and at $1/d_{w}\approx0.1$, say, after $t_{{\rm max}}$
updates a walker has typically only explored at most $l_{{\rm max}}\approx\left(10^{8}\right)^{0.1}\lesssim10$
sites from the origin; far too small a range to approximate asymptotic
scaling, especially in a heterogeneous structure such as HN3. 

\section{Renormalization Group Analysis\label{sec:Renormalization-Group-Analysis}}

To develop the renormalization group analysis for this problem, we
start from the master equation for the probability to be at site $n$
after $t$ update steps, $P_{n,t}$ \cite{Redner01}. In Fig. \ref{fig:4site},
we show an example set of recursions for some 5-site segment of HN3.
These on-site probabilities at subsequent times depend in delicate
ways on the inflow of probability from neighboring sites via the specific
local structure of incoming and outgoing long-rang links that can
vary a lot from segment to segment. 

In order to have a finite set of closed equations to obtain a closed
set of RG-recursions for hopping parameters, we study a sequence of
finite-sized systems with $N_{l}=2^{l}$ sites for $l=2,3,\ldots$.
For each, we use a generating function, 
\begin{eqnarray}
\tilde{P}_{n}(z)=\sum_{t=0}^{\infty}P_{n,t}z^{t},\label{eq:GenFcn}
\end{eqnarray}
in order to eliminate time dependence with this Laplace transform.
For the resulting hierarchy of ordinary recursion for a given $l$,
we then trace out every second site amplitude $\tilde{P}_{n}$, those
for odd values of $n$, by algebraic means. This obtains a new such
hierarchy in a form corresponding to a system corresponding to $l-1$
but with more complex hopping parameters. We repeat this procedure
until we discover a closed set of recursions amongst the hopping parameters.

\begin{figure}
\centering \includegraphics[bb=200bp 100bp 450bp 700bp,clip,width=0.8\columnwidth]{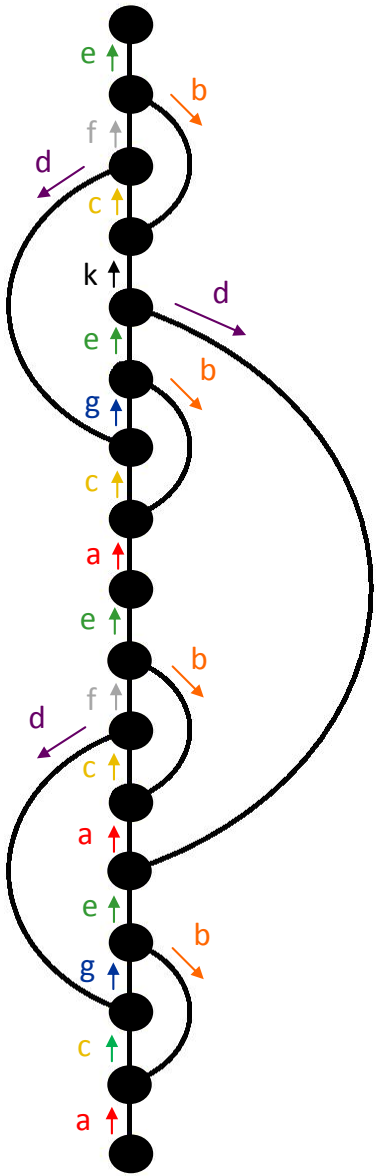}
\protect\caption{\label{fig:16color}Definition of renormalizable hopping parameters
for the biased walk along HN3. During each RG-step, every second site
is eliminated algebraically and a new set of equations result which
are identical in form to the previous set. Comparing the hopping parameters
in these equations before and after each step leads to the RG-flow
in Eq. (\ref{eq:RGflow}). }
\end{figure}

Here we demonstrate the RG process used to determine the recursion
equations by starting on a 16-site network ($l=4$) depicted in Fig.
\ref{fig:16color} and renormalizing it to an 8-site network ($l=3$),
revealing a self-similar pattern amongst the hopping parameters. First,
we rewrite the generic master equations, $P_{n,t+1}=\sum_{m}U_{n,m}P_{m,t}$,
as $\tilde{P}_{n}=z\sum_{m}U_{n,m}\tilde{P}_{m}$ by applying Eq.
(\ref{eq:GenFcn}), for some evolution matrix $U_{n,m}$ that describes
the hopping between sites. For the specific choice of $U_{n,m}$ refering
to the structure shown in Fig. \ref{fig:16color}, we reach the following
hierarchy of equations in terms of $z$ and $p$: 

\begin{eqnarray}
\tilde{P}_{0} & = & 1,\nonumber \\
\tilde{P}_{1} & = & z\tilde{P}_{0}+zp\tilde{P}_{3},\nonumber \\
\tilde{P}_{2} & = & z\tilde{P}_{1}+zp\tilde{P}_{6},\nonumber \\
\tilde{P}_{3} & = & z\tilde{P}_{2},\nonumber \\
\tilde{P}_{4} & = & z(1-p)\tilde{P}_{3}+zp\tilde{P}_{12},\nonumber \\
\tilde{P}_{5} & = & z\tilde{P}_{4}+zp\tilde{P}_{7},\nonumber \\
\tilde{P}_{6} & = & z\tilde{P}_{5},\nonumber \\
\tilde{P}_{7} & = & z(1-p)\tilde{P}_{6},\nonumber \\
\tilde{P}_{8} & = & z(1-p)\tilde{P}_{7}+zp\tilde{P}_{16},\label{eq:hierarchy16}\\
\tilde{P}_{9} & = & z\tilde{P}_{8}+zp\tilde{P}_{11},\nonumber \\
\tilde{P}_{10} & = & z\tilde{P}_{9}+zp\tilde{P}_{14},\nonumber \\
\tilde{P}_{11} & = & z\tilde{P}_{10},\nonumber \\
\tilde{P}_{12} & = & z(1-p)\tilde{P}_{11},\nonumber \\
\tilde{P}_{13} & = & z(1-p)\tilde{P}_{12}+zp\tilde{P}_{15},\nonumber \\
\tilde{P}_{14} & = & z\tilde{P}_{13},\nonumber \\
\tilde{P}_{15} & = & z(1-p)\tilde{P}_{14},\nonumber \\
\tilde{P}_{16} & = & z(1-p)\tilde{P}_{15}.\nonumber 
\end{eqnarray}
Here, we choose to terminated the lattice by connecting the \emph{8th}
site to the \emph{16th} site, instead of the \emph{24th} site as it
would have otherwise been by the rules of construction of the network
in Sec. \ref{sec:Network-Design}. This small change does not affect
the asymptotic behavior of the system at large sizes and times for
observables such as the MSD. 

After some trial-and-error, starting with a larger set of generalized
hopping parameters, we arrive at a minimal set necessary to parameterize
Eq. (\ref{eq:hierarchy16}) such that a closed system of RG-recursions
is obtained between those parameters. Inserting these parameters in
Eqs. (\ref{eq:hierarchy16}) yields 
\begin{eqnarray}
\tilde{P}_{0} & = & 1\nonumber \\
\tilde{P}_{1} & = & a\tilde{P}_{0}+b\tilde{P}_{3},\nonumber \\
\tilde{P}_{2} & = & c\tilde{P}_{1}+d\tilde{P}_{6},\nonumber \\
\tilde{P}_{3} & = & g\tilde{P}_{2},\nonumber \\
\tilde{P}_{4} & = & e\tilde{P}_{3}+d\tilde{P}_{12},\nonumber \\
\tilde{P}_{5} & = & a\tilde{P}_{4}+b\tilde{P}_{7},\nonumber \\
\tilde{P}_{6} & = & c\tilde{P}_{5},\nonumber \\
\tilde{P}_{7} & = & f\tilde{P}_{6},\nonumber \\
\tilde{P}_{8} & = & e\tilde{P}_{7}+d\tilde{P}_{16},\label{eq:16parametrized}\\
\tilde{P}_{9} & = & a\tilde{P}_{8}+b\tilde{P}_{11},\nonumber \\
\tilde{P}_{10} & = & c\tilde{P}_{9}+d\tilde{P}_{14},\nonumber \\
\tilde{P}_{11} & = & g\tilde{P}_{10},\nonumber \\
\tilde{P}_{12} & = & e\tilde{P}_{11},\nonumber \\
\tilde{P}_{13} & = & k\tilde{P}_{12}+b\tilde{P}_{15},\nonumber \\
\tilde{P}_{14} & = & c\tilde{P}_{13},\nonumber \\
\tilde{P}_{15} & = & f\tilde{P}_{14},\nonumber \\
\tilde{P}_{16} & = & e\tilde{P}_{15},\nonumber 
\end{eqnarray}
where we can read off the initial conditions from Eqs. (\ref{eq:hierarchy16})
for these generalized hopping parameters at RG-step $l=0$: 
\begin{eqnarray}
 &  & \begin{array}{lll}
a_{0}=z, & \quad b_{0}=zp, & \quad c_{0}=z,\\
\\
d_{0}=zp, & \quad e_{0}=z(1-p), & \quad f_{0}=z(1-p),\\
\\
g_{0}=z, & \quad k_{0}=z(1-p).
\end{array}\label{eq:RG-IC}
\end{eqnarray}
The parameters used in this system of equations are depicted in Fig.
\ref{fig:16color}.

A single step of the RG involves solving the hierarchy in Eq. (\ref{eq:16parametrized})
for $\tilde{P}_{n}$ with odd values of $n$, and then eliminating
them from every other equation, leaving equations that only contain
amplitudes with even $n$: 
\begin{eqnarray}
\tilde{P}_{0} & = & 1,\nonumber \\
\tilde{P}_{2} & = & \frac{ac}{1-bcg}\,\tilde{P}_{0}+\frac{d}{1-bcg}\,\tilde{P}_{6},\nonumber \\
\tilde{P}_{4} & = & eg\,\tilde{P}_{2}+d\,\tilde{P}_{12},\nonumber \\
\tilde{P}_{6} & = & \frac{ac}{1-bcf}\,\tilde{P}_{4},\nonumber \\
\tilde{P}_{8} & = & ef\,\tilde{P}_{6}+d\,\tilde{P}_{16},\label{eq:hierarchy8}\\
\tilde{P}_{10} & = & \frac{ac}{1-bcg}\,\tilde{P}_{8}+\frac{d}{1-bcg}\,\tilde{P}_{14},\nonumber \\
\tilde{P}_{12} & = & eg\,\tilde{P}_{10},\nonumber \\
\tilde{P}_{14} & = & \frac{ck}{1-bcf}\,\tilde{P}_{12},\nonumber \\
\tilde{P}_{16} & = & ef\,\tilde{P}_{14}.\nonumber 
\end{eqnarray}
Comparing the coefficients of these equations originating at the previous
RG-step to the corresponding coefficients of Eqs. (\ref{eq:16parametrized})
at the next RG-step gives us the recursions describing the RG-flow
of the system:
\begin{eqnarray}
a_{l+1} & = & \frac{a_{l}c_{l}}{1-b_{l}c_{l}g_{l}},\nonumber \\
b_{l+1} & = & \frac{zp}{1-b_{l}c_{l}g_{l}},\nonumber \\
c_{l+1} & = & e_{l}g_{l},\nonumber \\
e_{l+1} & = & e_{l}f_{l},\label{eq:RGflow}\\
f_{l+1} & = & \frac{k_{l}c_{l}}{1-b_{l}c_{l}f_{l}},\nonumber \\
g_{l+1} & = & \frac{a_{l}c_{l}}{1-b_{l}c_{l}f_{l}},\nonumber \\
k_{l+1} & = & \frac{k_{l}c_{l}}{1-b_{l}c_{l}g_{l}}.\nonumber 
\end{eqnarray}
The RG-flow in Eqs. (\ref{eq:RGflow}) describe how the generalized
hopping parameters transform under rescaling the system size from
$N$ to $2N$. At this point, we can further simplify these recursions.
First, note that there is no equation for the flow of the parameter
$d$ --- it maintains its initial value $d_{l}=d_{0}=zp$ throughout.
In this way, the RG-flow remains explicitly dependent on the microscopic
probability $p$, which will lead to the non-universality of the MSD
exponent in Eq. (\ref{eq:dw}). This explicit dependence of the RG-flow
on the control parameter is typical for hyperbolic networks that leads
to a number of novel critical phenomena \cite{Boettcher10c,Boettcher11d,BoBr12,PhysRevLett.108.255703}
in equilibrium thermodynamics. Secondly, we observe that the recursions
for $a_{l}$ and $k_{l}$, and consequently also for $f_{l}$ and
$g_{l}$ as well as $e_{l}$ and $c_{l}$, are identical, with the
distinction between each pair merely due to their difference in the
initial conditions in Eqs. \eqref{eq:RG-IC}. However, setting throughout
$k_{l}=\left(1-p\right)a_{l}$, $f_{l}=\left(1-p\right)g_{l}$, and
$e_{l}=\left(1-p\right)c_{l}$, which clearly reflects their corresponding
yet asymmetric roles at opposite ends of a long-range jump in Fig.
\ref{fig:16color}, leads to a much reduced set of equations:
\begin{eqnarray}
a_{l+1} & = & \frac{a_{l}c_{l}}{1-b_{l}c_{l}g_{l}},\nonumber \\
b_{l+1} & = & \frac{zp}{1-b_{l}c_{l}g_{l}},\nonumber \\
c_{l+1} & = & \left(1-p\right)c_{l}g_{l},\nonumber \\
g_{l+1} & = & \frac{a_{l}c_{l}}{1-\left(1-p\right)b_{l}c_{l}g_{l}},\label{eq:RGflow-1}
\end{eqnarray}
To verify the closure of the RG-flow, we consider a system of 64 sites
and confirm that the recursions maintain their self-similarity at
each RG-step. We conclude that the four distinct parameters given
in Eq. \eqref{eq:RGflow-1} are complete to describe the behavior
of the system \cite{Redner01}.

We note that this system of non-linear recursions, in fact, has exact
solutions for the case that $z=1$. Evolving the recursions from the
initial conditions, we find the relations $a_{l}c_{l}\equiv1$ and
$b_{l}\equiv pa_{l}$ to persist for all $l$. With that Ansatz, we
can reduce the system to a single recursion,
\begin{equation}
g_{l+1}=\frac{1}{1-p(1-p)g_{l}},\label{eq:gl-rec}
\end{equation}
which is similar to those found for continued fractions \cite{BO}.
It has the solution,
\begin{equation}
g_{l}=\frac{\left(\frac{1-p}{p}\right)^{l+1}-1}{p\left[\left(\frac{1-p}{p}\right)^{l+2}-1\right]},\label{eq:gl-sol}
\end{equation}
for $g_{0}=1$, from which the solutions for the other parameters
easily follows. Unfortunately, this solution can not be easily extended
to values of $z<1$. 

We now restrict our study to the fixed point (FP) of the RG-flow in
Eqs. \eqref{eq:RGflow-1} that describes the asymptotic properties
of the walk: For asymptotically large systems $l\to\infty$ we look
for stationary solutions setting $l\sim l+1$ to get 
\begin{eqnarray}
a_{\infty} & = & \frac{1-pz}{1-\left(1+z\right)p},\nonumber \\
b_{\infty} & = & \frac{(1-p)pz}{1-\left(1+z\right)p},\nonumber \\
c_{\infty} & = & \frac{1-\left(1+z\right)p}{1-p},\nonumber \\
g_{\infty} & = & \frac{1}{1-p}.\label{eq:FixedPoint}
\end{eqnarray}
To characterizes the dynamics of the system in the limit of large
systems at long times, we endeavor to study the simultaneous limits
of $l\to\infty$ and $\epsilon=1-z\to0$, to which end we linearize
in $\epsilon$ the RG-flow in Eqs. (\ref{eq:RGflow-1}) near the fixed
point. We do this by taking the Jacobian $J$ of these recursion equations,
which is the matrix consisting of the first-order derivatives of each
recursion equations with respect to each independent hopping parameter,

\begin{equation}
J_{l}=\left(\begin{array}{cccc}
\frac{c}{1-bcg} & \frac{ac^{2}g}{(1-bcg)^{2}} & \frac{a}{(1-bcg)^{2}} & \frac{abc^{2}}{(1-bcg)^{2}}\\
0 & \frac{pcg}{(1-bcg)^{2}} & \frac{pbg}{(1-bcg)^{2}} & \frac{pbc}{(1-bcg)^{2}}\\
0 & 0 & (1-p)g & (1-p)c\\
\frac{c}{1-(1-p)bcg} & \frac{(1-p)ac^{2}g}{\left[1-(1-p)bcg\right]^{2}} & \frac{a}{\left[1-(1-p)bcg\right]^{2}} & \frac{(1-p)abc^{2}}{\left[1-(1-p)bcg\right]^{2}}
\end{array}\right),\label{eq:J_l}
\end{equation}
which evaluated at the fixed point in Eqs. \eqref{eq:FixedPoint}
gives:

\begin{equation}
J_{\infty}=\left(\begin{array}{cccc}
1 & \frac{1}{1-2p} & \frac{(1-p)^{3}}{(1-2p)^{3}} & \frac{p(1-p)^{2}}{(1-2p)^{2}}\\
0 & \frac{p}{1-2p} & \frac{(1-p)^{2}p^{2}}{(1-2p)^{3}} & \frac{(1-p)^{2}p^{2}}{(1-2p)^{2}}\\
0 & 0 & 1 & 1-2p\\
\frac{1-2p}{(1-p)^{2}} & \frac{1-2p}{(1-p)^{3}} & \frac{1}{(1-p)(1-2p)} & \frac{p}{1-p}
\end{array}\right).\label{eq:Jinfty}
\end{equation}
Since by Eq. \eqref{eq:GenFcn} it is $z^{t}\sim e^{-\epsilon t}$,
the largest eigenvalue $\lambda$ of the Jacobian $J_{\infty}$ describes
the rescaling of time, $T\to T^{\prime}=\lambda T$ under rescaling
space, $L\to L^{\prime}=2L$, during the RG-step $l\to l+1$. The
largest eigenvalue of $J_{\infty}$ in Eq. \eqref{eq:Jinfty} is 
\begin{equation}
\lambda=\frac{2-3p}{1-2p}.\label{eq:lambda_max}
\end{equation}
The scaling Ansatz for the similarity scaling relation between time
and space, $T\sim L^{d_{w}}$, the yields $d_{w}=\log_{2}\lambda$,
for which the eigenvalue in Eq. \eqref{eq:lambda_max} finally gives
Eq. \eqref{eq:dw}.

\section{Conclusions\label{sec:Conclusions}}

We have presented a simple and exactly solvable example of a complex
(hyperbolic) network for which a biased random work leads to a non-universal
scaling behavior in which scaling exponents remain dependent on the
control parameter $p$ for the strength of the bias for $0\leq p\leq\frac{1}{2}$.
For $p>\frac{1}{2}$, the walker will stay confined near the origin,
because long-range jumps backwards along the network will be taken
frequently. Our analytical result, obtained via a renormalization
group calculation, are consistent with our numerical simulations for
$p<\frac{1}{2}$, and the scaling Ansatz fails for the data at $p>\frac{1}{2}$,
consistent with the exact result again. The origin of the non-universality
that is introduced through the non-renormalization of long-range links,
is reminiscent of similar findings for these hyperbolic networks in
equilibrium critical phenomena where they lead to novel synthetic
phase transitions that allow a high-level of engineering control \cite{Boettcher11d}.
While easy to engineer, critical phenomena on such hierarchical structures
tend to be very fragile to any form of disorder, however, which is
an issue that should be analyzed in more detail in the future.

\section*{Acknowledgements}

LAB thanks the Clare Boothe Luce Foundation for a fellowship. SB acknowledges
financial support from the U. S. National
Science Foundation through grant DMR-1207431. 

\bibliography{/Users/stb/Boettcher}

\end{document}